\documentclass[10pt, conference]{IEEEtran}

  	\usepackage[pdftex]{graphicx}
  	\graphicspath{{../pdf/}{../jpeg/}}
	\DeclareGraphicsExtensions{.pdf,.jpeg,.png}

	\usepackage[cmex10]{amsmath}
	\usepackage{algorithmic}
	\usepackage{array}
	\usepackage{mdwmath}
	\usepackage{mdwtab}
	\usepackage{eqparbox}
	\usepackage{url}
    \usepackage{amsfonts}
    \usepackage{graphicx}
    \usepackage{xcolor}
\begin{document}

\title{\LARGE Constrained Reinforcement Learning for Adaptive Controller Synchronization in Distributed SDN}

%DRL-driven Adaptive SDN Synchronization with contraint awareness.  

% \author{\authorblockN{Leave Author List blank for your IMS2013 Summary (initial) submission.\\ IMS2013 will be rigorously enforcing the new double-blind reviewing requirements.}
% \authorblockA{\authorrefmark{1}Leave Affiliation List blank for your Summary (initial) submission}}

\author{
\IEEEauthorblockN{Ioannis Panitsas, Akrit Mudvari, Leandros Tassiulas}
\IEEEauthorblockA{
Yale University, New Haven, CT\\
Emails: \{ioannis.panitsas, akrit.mudvari, leandros.tassiulas\}@yale.edu}
}
\maketitle

\begin{abstract}
In software-defined networking (SDN), the implementation of distributed SDN controllers, with each controller responsible for managing a specific sub-network or domain, plays a critical role in achieving a balance between centralized control, scalability, reliability, and network efficiency. These controllers must be synchronized to maintain a logically centralized view of the entire network. While there are various approaches for synchronizing distributed SDN controllers, most tend to prioritize goals such as optimization of communication latency or load balancing, often neglecting to address both the aspects simultaneously. This limitation becomes particularly significant when considering applications like Augmented and Virtual Reality (AR/VR), which demand constrained network latencies and substantial computational resources. Additionally, many existing studies in this field predominantly rely on value-based reinforcement learning (RL) methods, overlooking the potential advantages offered by state-of-the-art policy-based RL algorithms. To bridge this gap, our work focuses on examining deep reinforcement learning (DRL) techniques, encompassing both value-based and policy-based methods, to guarantee an upper latency threshold for AR/VR task offloading within SDN environments, while selecting the most cost-effective servers for AR/VR task offloading. Our evaluation results indicate that while value-based methods excel in optimizing individual network metrics such as latency or load balancing, policy-based approaches exhibit greater robustness in adapting to sudden network changes or reconfiguration. 
\end{abstract}

\IEEEoverridecommandlockouts

\IEEEpeerreviewmaketitle

\section{Introduction}
\subsection{Motivation}
Software-Defined Networking (SDN) \cite{6994333}, \cite{6739370} introduces a groundbreaking shift in network architecture, fundamentally differentiating the decision-making entity, known as the control plane, from the physical elements that handle traffic routing and forwarding, called the data plane. This separation enhances network performance through programmable management and reconfiguration, offering network operators increased flexibility and adaptability. SDN represents a major advancement in network design and management, providing enhanced efficiency and customization. In this architecture, a central SDN controller, which possesses a comprehensive understanding of the network's state, is responsible for distributing flow tables to the data plane devices for data routing. However, this centralized approach introduces challenges in scalability, availability, and security, making the SDN controller a focal point for potential attacks \cite{7150550} and a single point of failure. Due to its centralized nature, the controller struggles to efficiently manage network resources in large-scale environments, inherently limiting its scalability. Moreover, the centralization intensifies security vulnerabilities, as it presents a target for attackers, especially through DDoS attacks \cite{7275389}, which can significantly disrupt network availability and reliability. To address these challenges, SDN employs a distributed model \cite{8187644} with multiple independent controllers spread across the network. Although physically separated, these controllers operate in a 'logically-centralized' manner. This distributed SDN model, proposed to balance centralized and distributed control systems, involves each controller managing a specific set of subnetworks or domains. These controllers synchronize to maintain a cohesive, logically centralized network view. In large-scale networks, maintaining complete synchronization among controllers can be cost-prohibitive. Consequently, many distributed SDN networks adopt partial inter-controller synchronization, accepting temporary inconsistencies in the controllers' network views \cite{8470166}. This is known as the 'eventual consistency' model, which effectively manages costs and complexities.

Augmented Reality (AR) and Virtual Reality (VR) technologies are swiftly becoming popular due to their exceptional capacity to seamlessly integrate computer-generated graphics with real-life environments. These applications, while offering immersive experiences, are both computationally intensive and acutely delay-sensitive \cite{9284282}, \cite{9881517}. This makes their operation on mobile or local devices often impractical, particularly with the concern of battery life. A  solution is to offload the most resource-intensive computational tasks of AR/VR applications to proximate edge servers \cite{10.1007/978-3-319-74439-1_15}. However, in mobile edge computing environments, this offloading process encounters substantial challenges in adhering to the stringent delay requirements crucial for an uninterrupted AR/VR experience \cite{8885537}. Incorporating these applications into SDN adds another layer of complexity. SDN's ability to dynamically manage network resources and optimize data flow can be pivotal in meeting the low latency demands of AR/VR applications. Maintaining stringent latency requirements in SDN-controlled networks poses challenges, particularly given the distributed topology of SDN controllers. Furthermore, in dynamic wireless networks, network parameters like link latency and edge server computational capacity are in constant flux. Controllers must dynamically synchronized, ensuring efficient path selection to servers while adhering to latency constraints. This demands a proactive synchronization protocol, capable of real-time adjustments to the network's evolving topology and performance metrics, ensuring consistent quality of service and latency adherence.   

Inspired by the remarkable achievements of RL across various fields, such as mastering the game of Go \cite{silver2016mastering}, optimizing energy efficiency in data centers, researchers have applied various RL techniques to tackle synchronization problems in Software-Defined Networking (SDN). By formulating the synchronization problem as a Markov Decision Process (MDP), RL algorithms can learn the optimal policy for synchronizing SDN controllers, ensuring that the latency and QoS demands of AR/VR applications are met. 

\subsection{Methodology and Contributions}

In this paper, we developed a policy for synchronizing the distributed SDN controllers, adhering to the eventual consistency model. Inspired by recent successes in DRL, we formulated our problem as a MDP and we employed DRL algorithms for approximating the optimal synchronization policy. We examined both value-based and policy-based DRL methods to assist controllers in learning close to optimal synchronization policies and we evaluated them in various distributed SDN environments, where we varied multiple factors such as network topology, number of connected devices in each domain, latency constraints, and operational costs associated with edge servers. Our primary objective was to reduce network operation costs by offloading tasks to edge servers  while adhering to specific latency constraints. A secondary objective of our study was to expand the DRL framework to encompass existing SDN applications, including shortest path routing. This extension aimed to leverage and test the capabilities of our DRL framework in optimizing these traditional SDN functionalities. Finally, our last objective in this paper was to examine whether value based or policy based DRL algorithms are more effective in dynamic SDN settings, particularly in developing a synchronization policy in dynamic environments.  

The contributions of this work can be summarized as follows:
\begin{enumerate}
    \item We tackle the challenge of developing a synchronization policy in SDN environments, aiming to simultaneously ensure latency constraints and minimize network operator costs by strategically offloading tasks to edge servers.
    \item We show that policy-based DRL methods are superior in learning and converging faster to optimal synchronization policies when undergoing abrupt and significant changes or network reconfigurations, compared to value-based methods.  We are the first to implement policy-based RL algorithms for synchronizing distributed SDN controllers in dynamic networks.
    \item We evaluate our DRL framework in additional SDN applications, including shortest path routing. Our results demonstrate that value-based methods surpass policy-based approaches in effectively determining the optimal number of network paths and server allocations.
    \item We evaluate the robustness and adaptability of our DRL framework in a range of network conditions. This included altering the number of data and control plane devices, edge servers, as well as varying background traffic and server operational costs.
\end{enumerate}

\section{Related Works}

Recent breakthroughs \cite{silver2016mastering}, \cite{mnih2015human} have sparked considerable interest in applying RL techniques to complex decision-making tasks. In the context of SDN, the authors in \cite{8761183} propose a novel approach by combining RL techniques with deep neural networks. They develop the "Deep-Q (DQ) Scheduler", an advanced controller synchronization policy. This scheduler significantly outperformed the anti-entropy algorithm used in the ONOS controller, particularly for inter-domain routing tasks. In this work \cite{10278580}, the researchers applied a deep reinforcement and transfer learning-based method. This method equipped controllers with an efficient policy for synchronization and maintaining a logically centralized view in network systems. The effectiveness of this approach has been demonstrated across various applications, including shortest path routing and load balancing. The authors in \cite{8737388} identified the optimal frequency for synchronizing network views between controller pairs. Their approach, focusing on two objectives - maximizing synchronized controller pairs and enhancing application performance affected by synchronization rates - showed considerable advantages over traditional methods that synchronize all controller pairs at a uniform rate. In this study \cite{8888034}, the authors applied RL techniques combined with deep neural networks (DNNs) to develop a nuanced, scalable controller synchronization policy named "Multi-Armed Cooperative Synchronization" (MACS). The goal of MACS is to maximize the performance improvements gained through controller synchronizations, showcasing an innovative strategy in SDN management. Finally, in \cite{mudvari2023joint} the authors applied a multi-objective DRL approach for controller synchronization and placement.

\section{System Description}
% In this section, firstly, the distributed SDN environment and the SDN applications of interest are briefly described. Secondly, the controller synchronization problem and the synchronization rate are defined. Then, the MDP formulation is discussed, along with methods to solve it by employing RL algorithms for approximating the optimal synchronization policy. Both value-based and policy-based approaches are employed.

In this section, we begin by providing an overview of the distributed SDN environment and introducing the controller synchronization problem. Following this introduction, we delve into the applications of interest that are pertinent to this context. Subsequently, we explore the formulation of the synchronization problem as a MDP, discussing how RL algorithms can be utilized to approximate the optimal synchronization policy. In our approach, we employ both value-based and policy-based methods to address the problem effectively.

\subsection{Distributed SDN environment}

The network architecture consists of distributed controllers, denoted as \(c\), and data plane devices, denoted as \(d\). A network domain contains a variety of data plane devices, such as hubs, switches, and routers, as well as computing resources, including edge servers for task offloading.  Edge servers are utilized to increase the computational capacity of the system, primarily to offload tasks from various network applications. Each edge server has an assigned cost variable, indicating the expense involved in offloading and completing a task on that server. Neighbouring domains are interconnected through several gateway nodes. These gateways play a critical role in directing data flow and facilitating communication both across different domains and within the same domain. Data plane devices forward network traffic to the appropriate destination based on their flow table, which is controlled and updated by a control plane device, such as a distributed SDN controller. Control plane devices manage all critical network and security functions, including updating and distributing flow table information to data plane devices and filtering malicious traffic, among others. Link capacities and server costs are highly dynamic and are evolving due to the background traffic generated by users and network applications. Figure 1 illustrates a distributed SDN environment, including both control and data plane devices.

\begin{figure}[ht]
\centering
\includegraphics[width=\columnwidth]{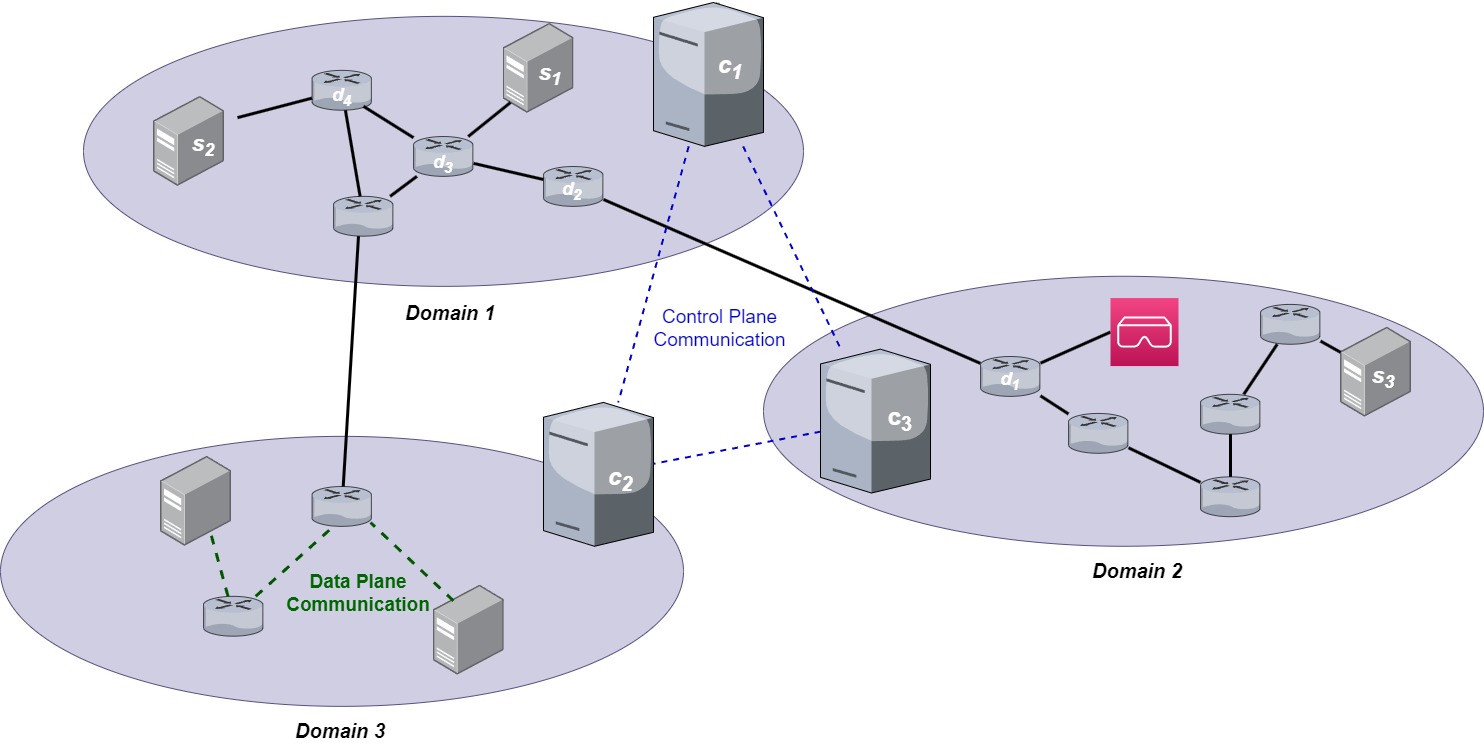}
\caption{A distributed SDN environment.}
\label{fig:network}
\end{figure}

\subsection{Controller Synchronization}

In this distributed SDN environment, our objective in this paper is to implement an efficient and effective controller synchronization algorithm under the eventual consistency model. Instead of synchronizing all the distributed controllers, we aim to synchronize only a subset of the available controllers while optimizing crucial metrics for different network applications. We define the subset of distributed controllers to synchronize in each time step as synchronization rate denoted as \(SR\). In each synchronization time step, after deciding which specific controllers to synchronize, network domain state information will be exchanged only with the synchronized controllers. This information can include link capacities, network domain load etc. By synchronizing only a subset of the total controllers, communication and signaling overheads will be minimized. The selection of controllers for synchronization needs to be carefully aligned with the goals of the network application. For instance, in the shortest path routing application, the main objective is to synchronize the most critical controllers that capture the dynamic nature of specific network domains, thereby maximizing the accurate number of the shortest paths between all source and destination nodes. This presents a significant challenge, especially when the synchronization rate is limited.

\subsection{Applications of Interest}

% In this paper, we primarily focused on developing a DRL framework for optimal task offloading in edge environments, with a particular emphasis on ensuring maximum latency thresholds while minimizing network operator costs. Although our concentration was mainly on AR/VR applications, our framework is developed for various applications requiring different constraints or aiming to optimize different objectives, such as selecting optimal servers with minimal network operator costs or minimal energy consumption. We selected AR/VR applications as our primary focus because they require substantial computing power and need to satisfy strict latency requirements for offloading tasks \cite{8319985}. In this context, we explored scenarios where users aim to offload AR/VR tasks to edge servers in domains that satisfy predefined and diverse latency criteria, while also seeking to minimize server-related expenses.

In this paper, our primary network application is offloading AR/VR tasks, that are generated by users in the network edge, to servers under pre-defined latency requirements. More specifically, recognizing the high computational demand and the low latency transmission requirements of AR/VR tasks \cite{8319985}, we considered an application where the generated tasks need to be transmitted under pre-defined latency constraints/requirements and concurrently the offloading procedure needs to be carefully designed to optimize network operator objectives. In our case, the objective was to minimize servers operation cost by selecting the optimal server for task offloading while adhering to pre-defined latency constraints. Latency requirements are divided into three categories: low latency task offloading, mid latency task offloading and high latency task offloading.
In the context of controller synchronization, the goal is to implement an algorithm or policy that strategically synchronizes a certain group of controllers based on the network application constraints and optimizing the network application objectives. This synchronization aims to achieve two main objectives: firstly, to ensure the maximum number of network paths meet the latency requirements for the tasks at hand, and secondly, to minimize network operator costs by efficiently allocating tasks to the most appropriate servers. Note that, the implemented policy can be easily modified and seamlessly integrated in every network application that follows the approach of optimizing network metrics under specific constraints. Figure 2, illustrates various scenarios encountered in the task offloading process for AR/VR applications. In the optimal scenario, tasks are offloaded to a server that both meets the latency requirements and minimizes computational costs. However, the figure also highlights suboptimal conditions, including cases where the latency constraints are not satisfied, the selected server does not provide the most cost-effective solution, or scenarios where neither latency nor cost criteria are adequately met.

Additionally, we evaluated our policy's effectiveness in a different network application, shortest path routing (SPR). In SPR, given the distributed architecture of controllers, each must determine global paths based on the network's logical view. However, due to the eventual consistency model, if a controller lacks accurate routing information, it may lead to incorrect path calculations \cite{kuzniar2015sdn}.

\subsection{MDP Formulation}
For approximating the optimal policy for our described network application, we formulated our problem as a MDP \cite{white2001markov}. MDP provides a mathematical framework ideal for modeling sequential decision-making scenarios. We use an MDP framework to model the decision-making process involved in synchronizing distributed network controllers.

\begin{itemize}
    % \item The state space $S$ is defined as a set within $\mathbb{R}^{1 \times N}$, where each state  \(s\in S\) is represented as a $1 \times N$ vector. State captures the cumulative time periods that have passed since the last update from each of the $N$ controllers. This representation enables us to monitor the frequency of updates from the controllers, providing insights into the synchronization policy and the network's dynamics.

    \item Each state \( s \) in state space \( S \) is represented as a vector \( s = [s_1, s_2, \ldots, s_n] \). Each element \( s_i \) in the vector corresponds to the synchronization status of the \( i^{th} \) SDN controller in a distributed environment comprising \( n \) controllers. The value of \( s_i \) indicates the number of time periods elapsed since the last synchronization of the \( i^{th} \) controller. 
    For instance, a state \( s = [0, 2, 0] \) in a system with three controllers (\( n = 3 \)) signifies that the first and third controllers were synchronized in the current time step (indicated by 0), while the second controller was last synchronized two time periods prior (indicated by 2). This representation enables us to monitor the frequency of updates from the controllers, providing insights into the synchronization policy and the network's dynamics.

    \item Each action \( a \) in action space \( A \) is represented as a one-dimensional binary vector \(a = [a_1, a_2, \ldots, a_n]\), where each component \(a_i\) corresponds to the decision made for the \(i^{th}\) controller. Specifically, \(a_i = 1\) indicates that the \(i^{th}\) controller is to be synchronized during the current time step, whereas \(a_i = 0\) signifies that the \(i^{th}\) controller will not be synchronized.
    Given a predefined synchronization rate \(SR\), the number of elements in the action vector \(a\) that are set to 1 must be exactly equal to \(SR\) following the eventual consistency model.

    \item  The reward function is an application-specific reward that depends on the primary objective of the application, as well as on whether or not the chosen policy violates certain constraint that the application may present. For instance, in the primary AR/VR application that we discuss, reward function is defined to impose penalties for synchronizing controllers that do not maximize the total number of correct network paths satisfying the latency requirements and penalizes decisions that do not minimize network operator costs constrained on the latency requirements. More specifically, the reward function is defined as follows:
        \begin{equation}
        R(s,a) = 
        \begin{cases} 
        -r_1, & \substack{\text{if } s_{\text{sel}} \in S \text{ and} \\ s_{\text{sel}} \notin S_L} \\
        -K (C(s_{\text{sel}}) - C(s_{\text{opt}})), & \substack{\text{if } s_{\text{sel}} \in S_L \text{ and} \\ C(s_{\text{sel}}) > C(s_{\text{opt}})} \\
        0, & \text{if } s_{\text{sel}} = s_{\text{opt}} \\
        -r_2, & \text{otherwise}
        \end{cases}
        \end{equation}
    
    Where:
    \begin{itemize}
        \item $s_{\text{sel}}$ is the edge server selected for task offloading.
        \item $S$ is the set of all available edge servers.
        \item $S_L$ is the subset of edge servers that meet the latency constraint $L$.
        \item $r_1$ and $r_2$ are negative rewards (penalties) for selecting a edge server that does not meet the latency or other criteria.
        \item $K$ is a scaling factor for the cost-based penalty.
        \item $C(s)$ is the cost associated with using server $s$.
        \item $s_{\text{opt}}$ is the optimal edge server that meets both latency and cost criteria.
    \end{itemize}

    For the non-constrained example, shortest path routing application, the reward function is defined as:
    \begin{equation}
    R(s,a) =  
    \sum_{t}k.\sum_{l}SP(l,t)
    \end{equation}
     where, $SP(l,t)$ is a binary variable in $\{0,1\}$ with value $1$ if the shortest path at time-step $t$ and link $l$ is accurately recognized, and $k$ is a scaling factor.

     More details about the selected values of the parameters can be seen in section IV-C.

    % \item The unconstrained reward function \(R(s,a)\) evaluates the immediate impact of taking action \(a\) in state \(s\) and it depends on the application goals and the specific scenarios it is intended to address. In our case, due to the constrained nature of our synchronization problem, we considered a reward function denoted as $\hat{R}(s,a)$ that incorporates the constraint parameters $ \lambda $ such as latency, energy, load balacing in the following way: 
    % \begin{equation}
    % \hat{R}(s,a) = f(R(s,a),\lambda))
    % \end{equation}
    % where, f is a function that takes us inputs the unconstrained rewards R and the constraint parameters $\lambda$ and returns the rewards $\hat{R}$ combining the constraints.
    % Intuitively, good controller synchronization actions under a specific SDN application will return higher returns and thus will help the agent to take better decisions in the future.
    % More details about the implemented reward functions for our applications of interest can be seen in the section IV-C.

\begin{figure*}
\centering
\includegraphics[width=\textwidth]{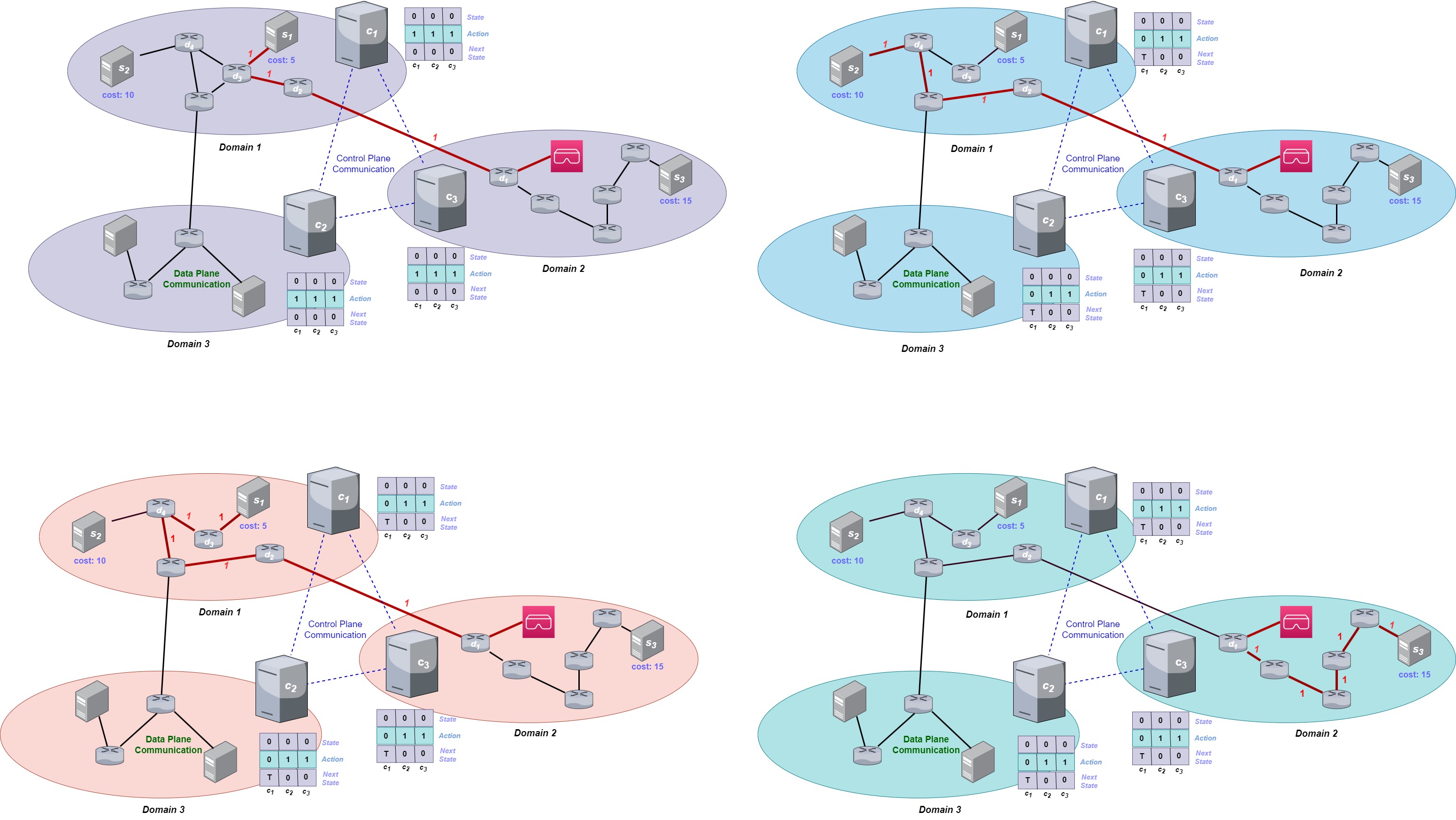}
\caption{Illustrating SDN controller synchronization with consistent and eventual consistency models. Top left: A consistent model where all controllers are synchronized, ideal for offloading an AR/VR task from domain 2 to an cost-effective edge server with maximum latency requirement equals to 4. Remaining images depict an eventual consistency model with a synchronization rate of 2. Top right: Meets latency requirements but not cost-effective. Bottom left: Optimal edge server selected for task offloading, but fails to meet latency constraint. Bottom right: Latency constraint and cost objective are unmet}
\label{fig:arvr_incostistency}
\end{figure*}
\end{itemize}

\subsection{Reinforcement learning formulation}

In RL, an agent takes sequential decisions by interacting with an environment, executing actions, and receiving feedback as rewards or penalties. The agent's goal is to develop a policy, a method for choosing actions based on its current state, to maximize the cumulative reward over time. This learning process is dynamic and adaptive, with the agent continually refining its policy based on ongoing interactions and feedback from the environment \cite{sutton2018reinforcement}. 
The primary aim of employing RL to our problem is to find the optimal policy denoted as $\pi^*$ for synchronizing the distributed controllers given the synchronization rate constraint \( SR \) for minimizing network operator costs, while adhering to one additional constraint: limiting latency to not exceed a pre-defined maximum value for an offloading task.

In the next sub-section, we will present value-based and policy-based RL methods, that applied to address the challenges in our synchronization problem. In this paper, we focused on a selection of value-based algorithms, specifically Deep Q-Networks and Double Deep Q-Networks. Additionally, we explored various policy-based methods, including the REINFORCE algorithm and Proximal Policy Optimization algorithm. While we also considered A2C-A3C, a hybrid approach combining aspects of both value and policy-based methods, it was excluded from our in-depth analysis due to its performance.

\subsection{Value Based Methods}

In value-based RL approaches, the primary objective is to estimate the value function, from which the policy is then inferred. These algorithms are categorized as off-policy, since the policy is determined based on the Q-function.  Value-based methods typically fall under the umbrella of model-free techniques. This implies that they operate without a predetermined model of the environment, learning to optimize the policy through direct interactions and experiences within the environment itself.

\noindent\subsubsection{Deep Q-Learning} 
\noindent Deep Q-Networks (DQNs) \cite{mnih2013playing}, developed by DeepMind, were a major breakthrough in RL, combining deep neural networks with Q-learning to handle complex, high-dimensional problems. The core concept employed a function approximator to estimate the Q-function using the Bellman equation for iterative updates. 
Inspired by this work, for a given synchronization state \(S\)  and rate \(SR\), our implemented Deep Q-network outputs a vector of action values \( Q(s, \cdot; \theta) \) to select which subset of controllers need to be synchronized at each time-step. The neural network's input comprises a state vector that indicates the number of time steps since each controller was last synchronized. The output of this neural network corresponds to the total number of possible actions, specifically detailing various synchronization methods for the controllers. We trained the Deep Q-network by minimizing a sequence of loss functions \( L_i(\theta_i) \) that change at each iteration \( i \). The loss function was defined as:
\begin{equation}
L_i(\theta_i) = \mathbb{E}_{s,a \sim \rho(\cdot)} \left[ (y_i - Q(s,a;\theta_i))^2 \right],
\end{equation}
where \( y_i = \mathbb{E}_{s' \sim E} \left[ r + \gamma \max_{a'} Q(s',a';\theta_{i-1}) | s, a \right] \) is the target for iteration \( i \), and \( \rho(s,a) \) is a probability distribution over sequences \( s \) and actions \( a \). The parameters from the previous iteration, \(\theta_{i-1}\), were held fixed when optimizing the loss function \( L_i(\theta_i) \). The derivative of the loss function was calculated as:
\begin{align}
\nabla_{\theta_i} L_i(\theta_i) = \mathbb{E}_{s,a \sim \rho(\cdot); s' \sim E} &\left[ \left( r + \gamma \max_{a'} Q(s',a';\theta_{i-1}) \right. \right. \nonumber \\
&\left. \left. - Q(s,a;\theta_i) \right) \nabla_{\theta_i} Q(s,a;\theta_i) \right],
\end{align}

\noindent Instead of computing the full expectations in this gradient, we utilized a replay memory, a data structure storing the agent's transitions, to approximate the expectation. By randomly sampling mini-batches from the replay memory, we were able to optimise the loss function through stochastic gradient descent
\begin{equation}
\theta_{i+1} \leftarrow \theta_i - \eta \nabla_{\theta_i} L_i(\theta_i),
\end{equation}
where \(\eta\) is the learning rate.

During the training phase, an epsilon-greedy strategy was employed. This method allowed the agent to explore and learn effective decision-making strategies by balancing between choosing random distributed controllers for synchronization (exploration) and the best-known controller synchronization action (exploitation).  Once training was completed, a greedy approach was adopted for selecting the optimal subset of controllers for synchronizing. This approach involved choosing the optimal controllers for synchronization with the highest Q-value from the deep Q-network's output layer, thereby identifying the optimal action for any given state.

\noindent\subsubsection{Double Deep Q-Learning} 
\noindent An issue encountered in the DQN algorithm was that the gradient descent updates of \(\theta\) differed from classical supervised learning. This distinction arose because the training target
\begin{equation}
y_i = R(s, a) + \gamma \max_{a'} Q_i(s', a'; \theta_i)
\end{equation}
was generated by the same Q-function \(Q_i(s, a; \theta_i)\) that was trained. This made it more likely to select overestimated values, resulting in overoptimistic value estimates. Double Q-learning (DDQN) \cite{van2016deep}, an updated version of DQN, prevented this issue by decoupling the selection from the evaluation. For that reason, we used two value functions that were learned by assigning each experience randomly to update one of the two value functions. This resulted in two sets of weights, \(\theta\) and \(\theta'\). For each update, one set of weights was used to determine the greedy action (optimal action), and the other set was used to determine its value. The secondary network learned at a slower pace compared to the main network, which helped to prevent overfitting. The updated training target value can be calculated as follows:
\begin{equation}
y_i = R(s, a) + \gamma \ Q_{\theta'}(s', \text{argmax}_{a'} Q_{\theta}(s', a'))
\end{equation}

 Additionally, we incorporated a replay memory mechanism in our training process to store training samples, allowing the agent to revisit previous experiences. Once training was completed, a greedy approach was adopted for controller selection. This approach involved choosing the optimal controllers for synchronization with the highest Q-value from the neural network's output layer.  Once training was completed, a greedy approach was adopted for selecting the optimal subset of controllers for synchronizing similar to DQN approach.

\subsection{Policy Based Methods}
In policy-based RL approaches, the strategy is to approximate a stochastic policy $\pi$ directly instead of deriving a deterministic policy from an estimated value function. A stochastic policy is a function $\pi : \mathcal{S} \rightarrow \mathcal{P}(\mathcal{A})$ that maps each state \(s\) to a probability distribution over actions assigning a probability \(\pi(a|s)\) for each action \(a\) in that state.
  A stochastic policy is often estimated with a function approximator having trainable parameters $\theta$. The policy is parameterized as \( p_{\theta}(a|s) \), where the primary objective involves computing the gradient of the expected rewards with respect to the policy parameters \( \theta \). Subsequently, gradient ascent is applied to iteratively refine \( \theta \), thereby enhancing the maximization of cumulative long-term rewards.

\noindent\subsubsection{REINFORCE} 
\noindent REINFORCE \cite{williams1992simple} is a policy gradient algorithm, that employs a neural network to approximate the stochastic policy, characterized by parameters \(\theta\). We implemented a neural network that takes as input the distributed controllers state and produces a probability distribution over possible pairs of controllers to synchronize. We directly computed the gradient of the long term expected return denoted as $\nabla_{\theta} J(\theta) $, with respect to the policy parameters \( \theta \) and we used gradient ascent for updating the trainable parameters. 

% \begin{equation}
% J(\theta) = \mathbb{E}_{\tau \sim \pi_\theta} \left[ \sum_{t=0}^{T} \gamma^t R(s, a) \right]
% \end{equation}

% \noindent where \( \mathbb{E}_{\tau \sim \pi_\theta} \) denotes the expected sum of rewards over trajectories \( \tau \) under the policy \( \pi \) parameterized by \( \theta \).

\noindent The gradient was calculated as follows:

\begin{equation}
\nabla_{\theta} J(\theta) = \mathbb{E}_{\tau \sim \pi_\theta} \left[ \sum_{t=0}^{T} \gamma^t R(s, a) \nabla_{\theta} \log \pi_\theta(a_t | s_t) \right]
\end{equation}

% \noindent In practice, computing this gradient exactly is often infeasible due to the complexity of the expectation over all possible trajectories. Therefore, we approximate the gradient by sampling. The approximate gradient is computed as:

% \begin{equation}
% \nabla_{\theta} J(\theta) \approx \frac{1}{N} \sum_{n=1}^{N} \sum_{t=0}^{T} \gamma^t R(s_t^n, a_t^n) \nabla_{\theta} \log \pi_\theta(a_t^n | s_t^n)
% \end{equation}

% where \( N \) is the number of trajectories sampled, and \( (s_t^n, a_t^n) \) represents the state and action at time \( t \) in the \( n \)-th trajectory. 
\noindent To optimize \( J(\theta) \) and improve the policy, the parameters \( \theta \) are updated using gradient ascent:

\begin{equation}
\theta \leftarrow \theta + \eta \nabla_{\theta} J(\theta)
\end{equation}

\noindent where, \( \eta \) is the learning rate. 
This process iteratively adjusts the policy towards higher expected returns by following the direction of the approximated gradient.
\noindent\subsubsection{Proximal Policy Optimization} 
\noindent 
The fundamental challenge with certain methodologies in policy optimization arises when implementing a large step size. Such an approach can lead to substantial alterations in the policy at each iteration. This significant change can result in a considerable gap between the parameters of the consecutive policies — the new and the old — which, in turn, risks leading to a collapse of the model. 
To address this issue, Proximal Policy Optimization (PPO) \cite{schulman2017proximal} emerges as an advanced strategy, refining the traditional policy gradient methods. PPO's core objective is to facilitate substantial updates in the policy while concurrently imposing a constraint. This constraint, known as the trust region constraint, ensures that the new and old policies remains within a predefined limit.  
The objective of PPO is encapsulated in the following equation, which defines its core operational mechanism:

\begin{equation}
\begin{aligned}
L^{CLIP}(\theta) = \hat{E}_t \bigg[ \min\bigg(&\frac{\pi_\theta(a_t|s_t)}{\pi_{\theta_{old}}(a_t|s_t)} \hat{A}_t, \\
&\text{clip}\left(r_t(\theta), 1 - \epsilon, 1 + \epsilon\right) \hat{A}_t\bigg) \bigg]
\end{aligned}
\end{equation}
where   $ r(\theta)$ denotes the probability ratio:
\begin{equation}
 r_t(\theta) = \frac{\pi_\theta(a_t|s_t)}{\pi_{\theta_{old}}(a_t|s_t)}.
\end{equation}

\noindent The first component of the objective function represents the core objective of the TRPO algorithm. However, this part of the function does not incorporate a constraint. As a result, maximizing this objective without any restrictions could potentially lead to unduly large updates in the policy, a scenario that might be undesirable in maintaining the stability and effectiveness of the algorithm.
By introducing the second term, $\text{clip}(r_t(\theta), 1 - \epsilon, 1 + \epsilon) \hat{A}_t$, PPO alters the surrogate objective by implementing a clipping mechanism on the probability ratio. This modification effectively constrains the expansion of \( r_t \) beyond the interval [1 - $\epsilon$, 1 + $\epsilon$]. Consequently, the objective is finalized by choosing the lesser of the clipped or unclipped objectives, thus establishing the ultimate objective as a conservative estimate, or a 'guarded bound', on the unclipped objective. This approach ensures that alterations in the probability ratio are taken into account only when they adversely affect the objective, fostering a more balanced policy update process.

In our PPO implementation, we utilized a neural network to directly approximate the synchronization policy. The input to this neural network is the controllers state information while the output is a probability distribution across all possible actions, where each action corresponds to a different method of synchronizing the controllers. This setup allows our agent to select the most optimal actions based on the generated probability distribution, thereby eliminating the need for an epsilon-greedy strategy due to the stochastic nature of the output layer. During training, the focus was on refining the weights of the neural network  in order to select optimal controllers that adhere to latency constraints and efficiently allocate edge servers, while minimizing the objective function $L^{CLIP}(\theta)$ using stochastic gradient descent.

More details about the selected hyper-parameters of the above algorithms can be seen in section IV-C.

\section{Evaluation}
This section begins by introducing the performance benchmarks in Section IV-A. Then, the distributed network settings, reinforcement learning settings, and experimental settings for evaluation scenarios are described in Sections IV-B, IV-C, and IV-D, respectively. Finally, the evaluation results and analysis are presented in Sections IV-E, IV-F, and IV-G.

\begin{figure}[ht]
\centering
\includegraphics[width=\columnwidth]{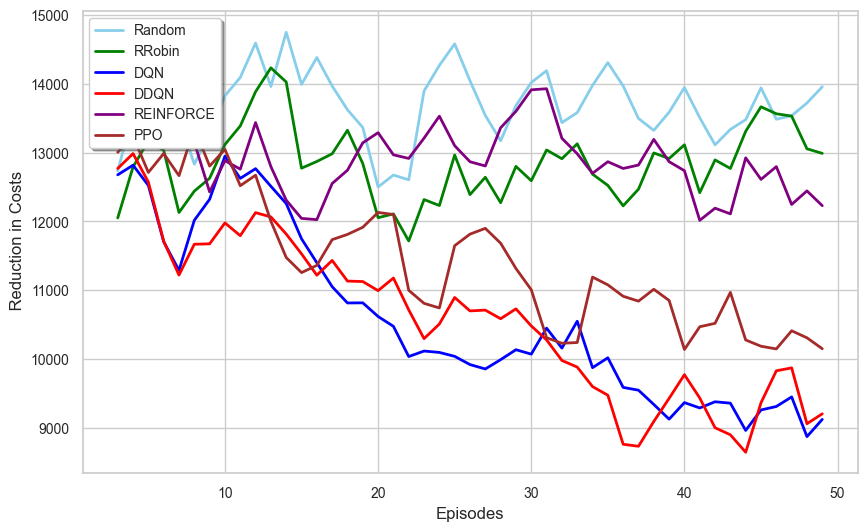}
\caption{Value-based algorithms outperform both policy-based and baseline strategies in terms of cost minimization.}
\label{fig:costs}
\end{figure}

% Compared to Random: 22.95% Compared to RRobin: 17.77% Compared to DDQN: 0.45% Compared to REINFORCE: 18.37%  Compared to PPO: 7.32%

\begin{figure}[ht]
\centering
\includegraphics[width=\columnwidth]{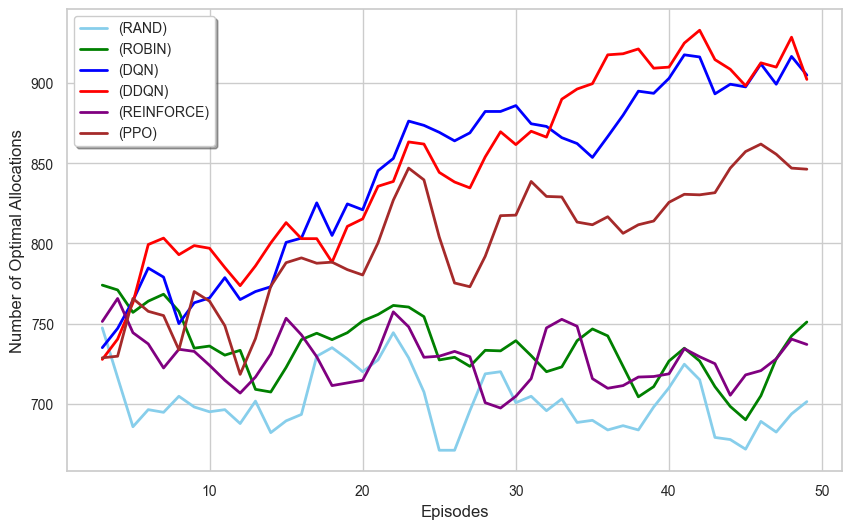}
\caption{Value-based algorithms outperform both policy-based and baseline strategies in terms of finding correct allocations.}
\label{fig:allocations}
\end{figure}

% Compared to RAND (Random): 20.61% Compared to ROBIN: 15.15% Compared to DQN: 0.28% Compared to REINFORCE: 16.41% Compared to PPO: 5.70%

\begin{figure}[ht]
\centering
\includegraphics[width=\columnwidth]{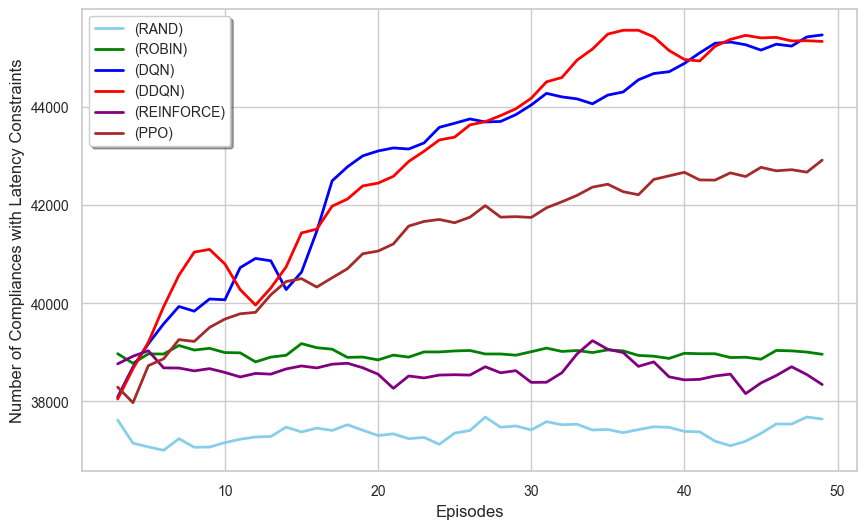}
\caption{Value-based algorithms outperform both policy-based and baseline strategies in finding correct path that satisfy latency constraints.}
\label{fig:paths}
\end{figure}

% Compared to RAND (Random): 15.18% Compared to ROBIN: 10.45% Compared to DQN: 0.19% Compared to REINFORCE: 11.53% Compared to PPO: 4.26%

\begin{figure*}[ht] 
\centering
\includegraphics[width=\textwidth]{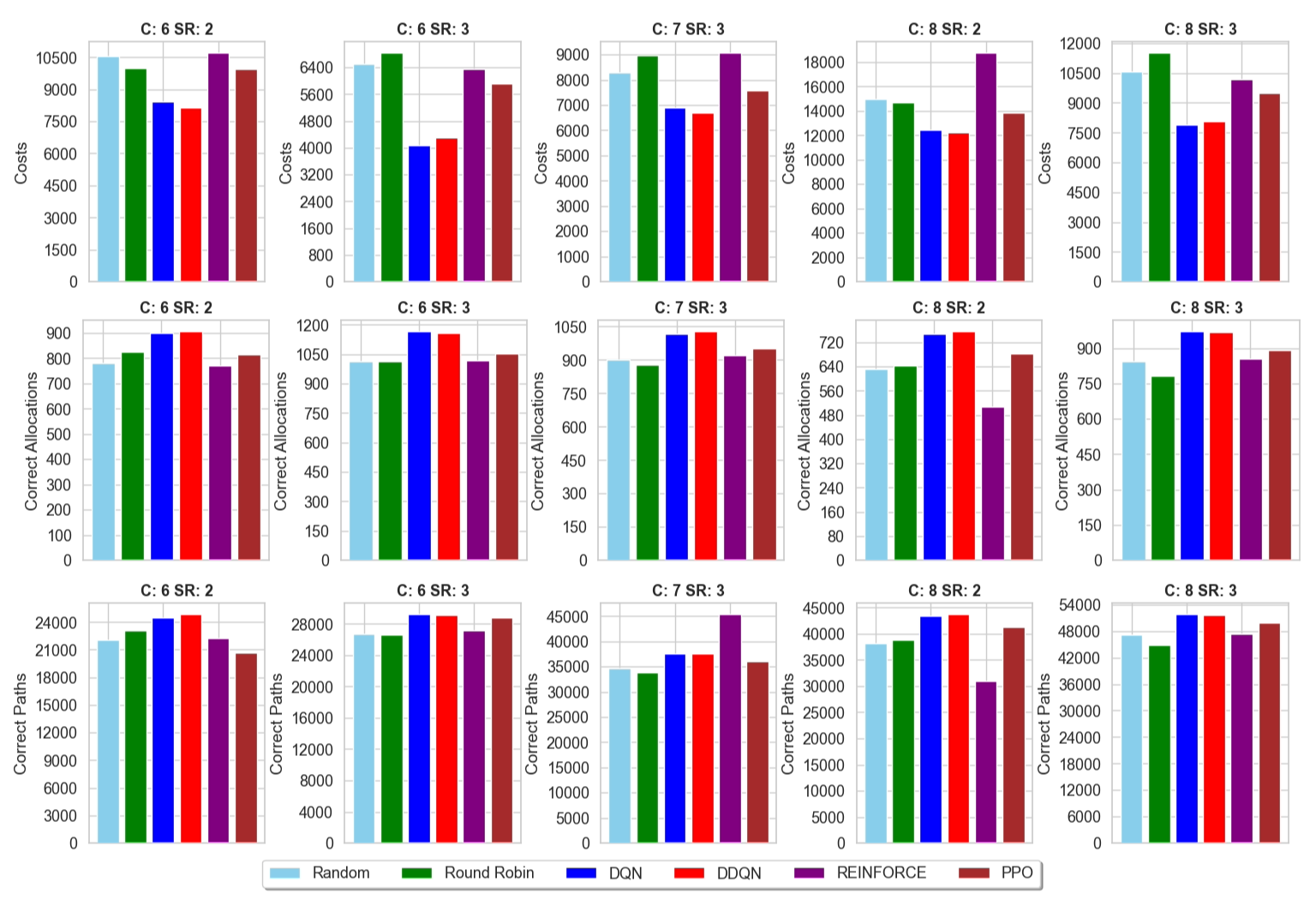} 
\caption{Costs, correct server allocations and network paths of value-based and policy-based algorithms as well as random and round-robin baselines for low latency task-offloading.}
\label{fig:1_latency}
\end{figure*}

\begin{figure*}[ht] 
\centering
\includegraphics[width=\textwidth]{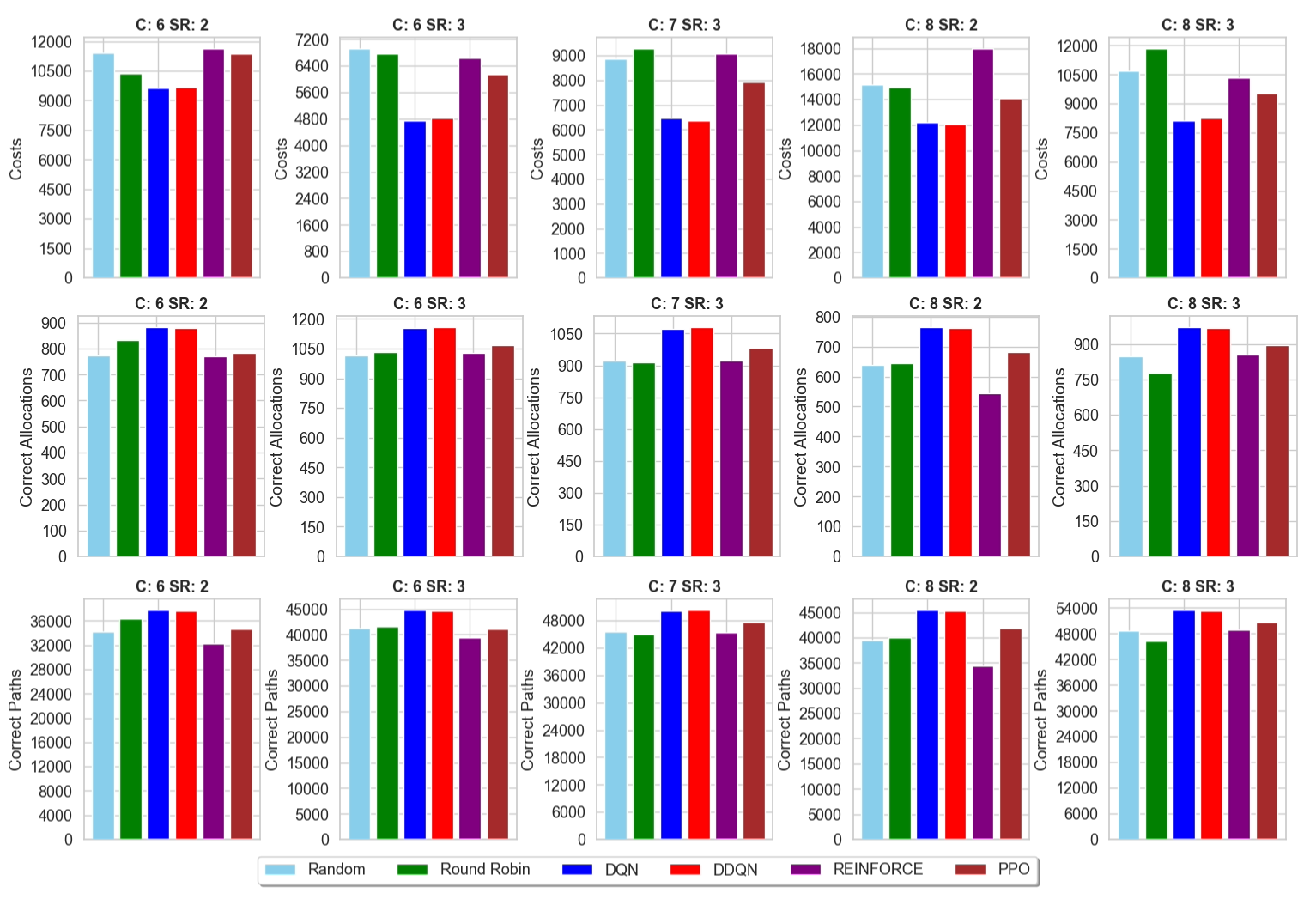} 
\caption{Costs, correct server allocations and network paths of value-based and policy-based algorithms as well as random and round-robin baselines for mid latency task-offloading.}
\label{fig:2_latency}
\end{figure*}

\begin{figure*}[ht] 
\centering
\includegraphics[width=\textwidth]{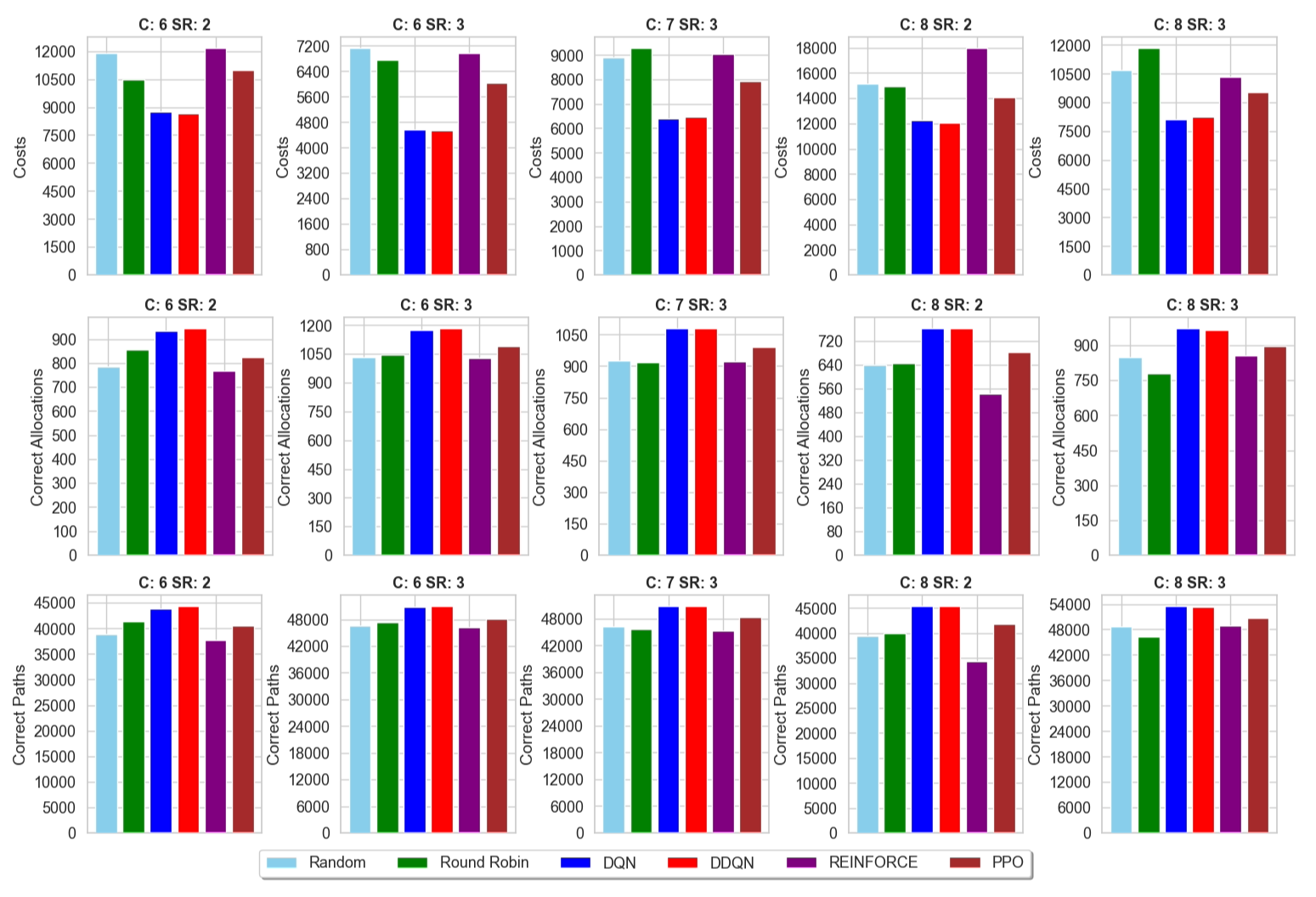} 
\caption{Costs, correct server allocations and network paths of value-based and policy-based algorithms as well as random and round-robin baselines for high latency task-offloading.}
\label{fig:3_latency}
\end{figure*}

\subsection{Performance Benchmarks}
 For bench-marking purposes, we specifically selected round robin (RRobin) and random synchronization methods for their straightforwardness and intuitive nature.
In round robin synchronization, domain controllers are sequentially synchronized, starting with a set of \( SR \) controllers in a cyclical pattern. Conversely, random synchronization selects \( SR \) controllers randomly at each time step for a varied sequence.

\subsection{Distributed SDN Environment Settings}
Our distributed network environment is composed of various domains and distributed controllers. Experiments were conducted with a variety of network configurations, altering the numbers of data plane devices, controllers, and edge servers. Each domain in our network is assigned a distributed controller, responsible for maintaining routing information and monitoring link statuses. Within each domain, there are multiple servers available for offloading tasks from AR/VR applications, with each server allocated a cost reflective of real-world scenarios. For this scenario, we generated server costs as uniformly random values between 20 and 100, denoted as \(C\). It's important to note that these values can be interpreted with a different minimization objective; rather than minimizing network costs, tasks are offloaded to edge servers that consume the least energy for the task. Additionally, we created traffic patterns that closely mimic reality, based on selected network metrics. The network was tested across 5 to 12 domains, with each domain comprising a random number of data plane devices ranging from 2 to 15. While we established a synchronization rate ranging from 2 to 4 for each network topology implemented, in section IV-E, we specifically present results using 6 to 8 network domains and synchronization rates ranging from 2 to 3.

\subsection{Reinforcement Learning Settings}

Our DRL framework was developed using the PyTorch library in Python. For the DQN and DDQN models, we designed a neural network architecture with a single hidden layer comprising 50 neurons and utilized ReLU as the activation function. The training parameters included a batch size of 256 and an epsilon decay factor of 10 to adjust the likelihood of random actions. The replay memory was set at a capacity of 40000. Throughout the training, we employed the Mean Squared Error (MSE) as the loss function and used the Adam optimizer with a learning rate of 0.01. Additionally, we set a discount factor of 0.1 in these models to better assess future rewards. Regarding policy gradient methods, specifically REINFORCE and PPO, we followed a similar neural network structure with a single hidden layer of 50 neurons using ReLU for activation. The output layer, however, utilized a Softmax layer for generating probability distributions. These models were trained with a batch size of 256, using the Adam optimizer at a learning rate of 0.001, and each training cycle included 5 iterations. For effective learning, we opted for a clip epsilon of 0.1 and a gradient clipping norm of 7. Like the DQN and DDQN models, these policy gradient methods also incorporated a discount factor of 0.1. 

For the AR/VR application, the parameters of the reward function were initiated with these values \( r_1 = -10000 \), \( r_2 = -8000 \), and \( K = 80 \). For the SPR application, the scalar variable \(k\) was set \(k=100\).

\subsection{Experimental Settings}

Our experiments were carried out on a high-performance computer equipped with an Intel(R) Core(TM) i9-10920X CPU running at 3.50GHz, paired with an NVIDIA GPU, specifically the GeForce RTX 3090.

\subsection{Evaluation Results}

\noindent Analysis of Figures 3, 4, and 5 shows that value-based algorithms perform better than policy-based algorithms, as well as the random and round robin synchronization algorithms, in terms of minimizing the network operator's costs, maximizing the number of optimal server allocations, and maximizing the number of correct network paths, respectively. PPO does seem to perform quite well as opposed to other non-value based methods, with results much closer to value based approaches. Specifically, Figure 3 highlights the efficiency of the DQN algorithm in reducing network operator costs. It shows that DQN outperforms other methods, being 0.45\% more effective than DDQN, 7.32\% more than PPO, 17.77\% more than Round Robin, 18.37\% more than REINFORCE, and 22.95\% more effective compared to the Random algorithm. Moreover, Figure 4 illustrates that the DDQN algorithm excels in identifying optimal server allocations. Specifically, it outperforms the DQN by 0.28\%,  PPO by 5.70\%, Round Robin by 15.15\%, REINFORCE by 16.41\% and Random by 20.61\%. In Figure 5, the results indicate that DDQN once again emerges as the optimal algorithm for maximizing the correct number of network paths within the given latency constraints. DDQN shows a 0.19\% advantage over DQN, 4.26\% over PPO, 10.45\% over Round Robin, 11.53\% over REINFORCE, and 15.18\% over Random.

Additionally, we evaluated our DRL framework across various network topologies, focusing on controllers ranging from 6 to 8 and synchronization rates between 2 and 3, as well as, different latency requirements. As depicted in Figure 6 (low latency, 8 ms), Figure 7 (mid latency, 10 ms), and Figure 8 (high latency, 12ms), we observed that value-based algorithms outperform both policy-based methods and traditional synchronization approaches like random and round-robin in terms of reducing network operator costs while also maximizing the efficiency in server allocations and identifying correct network paths. Specifically, for low latency task offloading scenarios (shown in Figure 6), we found that both DQN and DDQN exhibit similar levels of high performance, making them preferred choices for this application. Among the policy-based methods, PPO demonstrated notable effectiveness in reducing network costs and successfully finding correct servers and paths. In scenarios with mid latency requirements, as depicted in Figure 7, the DQN and DDQN algorithms show remarkably similar effectiveness, marking them as optimal choices for these conditions. The PPO algorithm also delivers a robust performance, although not quite on par with DQN and DDQN. In contrast, the REINFORCE algorithm, along with random and round robin approaches, demonstrate a lower level of performance, indicating their limited suitability for these specific latency demands. For high latency requirements as shown in Figure 8, DQN and DDQN again stand out as the most effective algorithms, with PPO also showing strong performance. However, REINFORCE, random, and round robin methods lag behind in effectiveness for these requirements.

\subsection{Additional Application: Shortest Path Routing}

To further validate our DRL framework beyond varying network environments and latency requirements for AR/VR task offloading, we extended our evaluation to a distinct SDN application: shortest path routing. Here, we examined only the best-performing algorithm from each category - value-based and policy-based. As illustrated in Figure 9, DDQN emerged as the most accurate in detecting correct network paths with an accuracy of 86.76\%, closely followed by PPO at 85.98\%. Comparatively, the random and round robin methods achieved lower accuracies of 84.6\% and 80.7\%, respectively.

\begin{figure}[ht]
\centering
\includegraphics[width=\columnwidth]{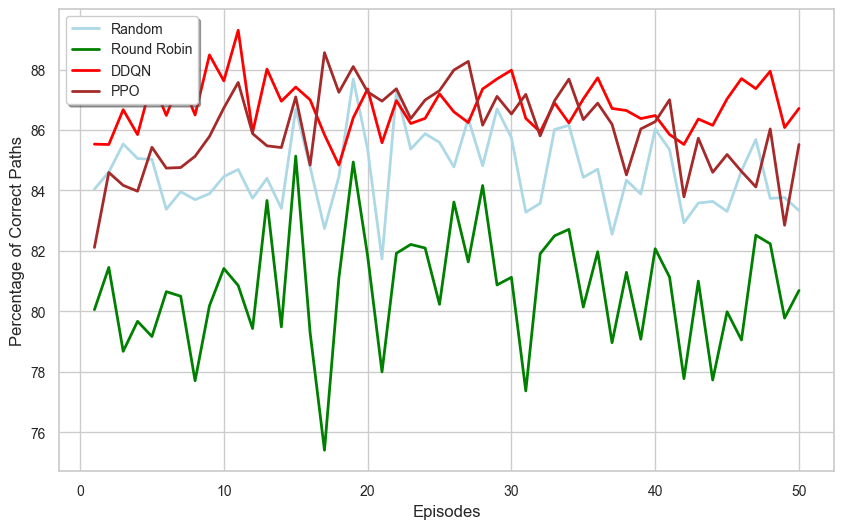}
\caption{Percentage accurate detection of shortest paths under different reinforcement learning algorithms}
\label{fig:spr1}
\end{figure}

%  Random    84.6%    RRobin    80.7%   DDQN    86.76%   PPO 85.98%

\subsection{Robustness to significant network changes or reconfiguration}

\begin{figure}[ht]
\centering
\includegraphics[width=\columnwidth]{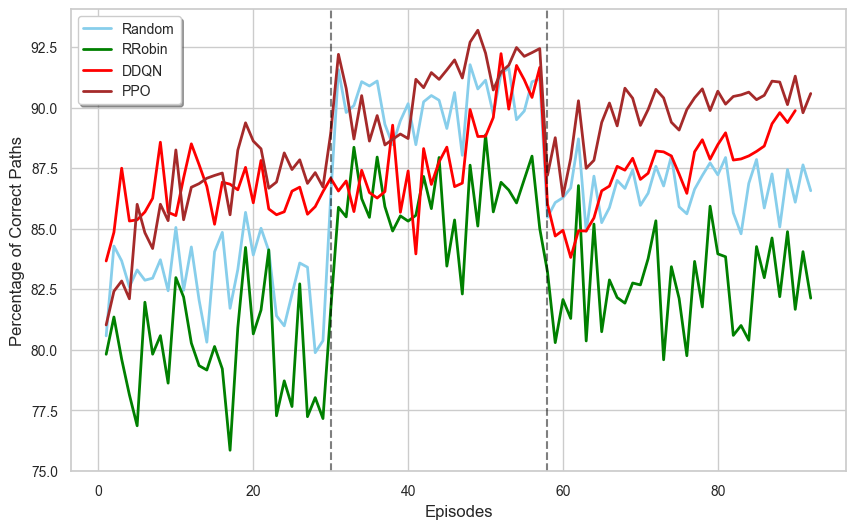}
\caption{Percentage accurate detection of shortest paths with network reconfiguration at random intervals}
\label{fig:reconfig}
\end{figure}

% Random    86.5%   RRobin    82.8%   ppo   89.0%    ddqn      87.23%

In this section, our goal will be to observe the effects of random and abrupt network reconfiguration (or significant changes to the network), which is a scenario where the subdomain (nodes/switches and links) assigned to each neighboring controller is changed. On top of the previous assumptions of a dynamic network, i.e., where network structure is changing relatively slowly, in this new paradigm the network will undergo significant reconfiguration (links and nodes are randomly recreated). As shown in Figure 10, the network undergoes such reconfiguration at episodes 30 and 58 (denoted by horizontal dotted line.). Like in previous cases, we see that the learning occurs better in RL methods, with value-based DDQN approaching the most efficient policy faster than PPO. But when reconfiguration happens, PPO is shown to be more robust to the significant network change occurring after episodes 30 and 58; it is shown to be capable of reaching higher efficiency in the new environment very fast after each. PPO outperforms with 89\% accuracy in path detection, closely followed by DDQN at 87.23\%, random at 86.5\%, and round robin at 82.8\% This implies that, at least in absence of further modifications to value based methods, which could be change in exploitation vs. exploration trade off among other improvements, PPO is a better alternative for extremely vocative network configurations. 

\section{Conclusion}

In this paper, we address the synchronization problem in distributed SDN controllers under constrained synchronization rates and latency requirements. Our goal was to create a DRL-based synchronization policy for efficient AR/VR application offloading, targeting both an upper limit on network delay and cost reduction for network operators. In addition to exploring value-based methods such as DQN and DDQN, our research was extended to include policy-based RL techniques, and their application in other network tasks such as shortest path routing. Initially, we formulated the synchronization problem as a MDP and subsequently defined our  SDN environment. Besides approximating the optimal synchronization policy with value-based RL algorithms, we also employed policy-based algorithms and explored their benefits. Furthermore, this research was not limited to developing strategies for optimizing an application objective under a strict constraint, but also included an evaluation of our RL framework across multiple network applications. Our evaluation results indicate that while value-based methods outperform in optimizing single network metrics such as latency followed closely by PPO, policy-based approaches are more robust in sudden network changes or re-configurations and can achieve higher performance in fast evolving dynamic networks.

% ==================
% # ACKNOLEDGMENTS #
% ==================

%\section*{Acknowledgment}

% ==============
% # REFERENCES #
% ==============

\bibliographystyle{IEEEtran}
\bibliography{IEEEabrv,biblio_traps_dynamics}

\end{document}